\documentclass[prc,aps,showpacs]{revtex4}
\usepackage{graphicx}
\begin{document}

\title{Critique of multinomial coefficients method for evaluating Tsallis and R\'enyi entropies}

\author{A.S.~Parvan$^{a,b}$}

\affiliation{$^{a}$Bogoliubov Laboratory of Theoretical Physics, Joint Institute for Nuclear Research, 141980 Dubna, Russian Federation}

\affiliation{$^{b}$Institute of Applied Physics, Moldova Academy of Sciences, MD-2028 Chisinau, Republic of Moldova}


\begin{abstract}
Oikonomou [Physica A 386 (2007) 119] has published a calculation which purports to show that the Tsallis and R\'enyi entropies can be obtained from the generalized multinomial coefficients. In this paper, we prove that the method of generalized multinomial coefficients failed to determine the Tsallis entropy at equilibrium. Moreover, it is shown that Oikonomou's analysis contains mistakes which led him to misleading statements related to the Jaynes
principle of maximum entropy, the Tsallis and the R\'enyi statistics.
\end{abstract}
\maketitle

The paper~\cite{Oikon} examines the ideal Maxwell-Boltzmann gas of $N$ identical particles in the microcanonical ensemble in the approximation of the degenerate single-particle states of a system without the Gibbs correcting term in the framework of the Boltzmann-Gibbs, Tsallis and R\'enyi statistical mechanics. In~\cite{Oikon}, the Tsallis and R\'enyi entropies for the ideal gas were directly determined from the thermodynamical relations of the ideal gas of the Boltzmann-Gibbs statistics by the generalization of its multinomial coefficients. In this paper, we derive the Tsallis and R\'enyi entropies for the ideal gas from their general definitions in the framework of the Tsallis statistics~\cite{Tsal88,Tsal98,Toral,Parv1,Parv2} and the R\'enyi statistics~\cite{Renyi,Wehrl,Lenzi00,ParvBiro,ParvBiro1} at equilibrium and compare them with the results obtained in~\cite{Oikon}. We will prove that the Tsallis entropy of the ideal gas obtained by Oikonomou in~\cite{Oikon} from the generalized multinomial coefficients is
incompatible with the entropy of the ideal gas of the self-consistent Tsallis statistics.

The Boltzmann-Gibbs entropy for the ideal Maxwell-Boltzmann gas of $N$ identical particles in the occupation number representation~\cite{Huang,Landau} can be written through the multinomial coefficients in the form~\cite{Oikon}
\begin{equation}\label{1}
    S_{BG} =-N\sum\limits_{i=1}^{W} p_{i} \ln p_{i}=
   \ln W\{\langle n_{i}\rangle\}, \qquad  p_{i}=\frac{\langle n_{i}\rangle}{N}
\end{equation}
and
\begin{eqnarray}\label{2}
    W\{n_{i}\}&=& \lim_{N\to\infty} C_{\{n_{i}\}}^{N} =
    \prod\limits_{i=1}^{W}
    \left(\frac{N}{n_{i}}\right)^{n_{i}}, \\ \label{3}
 C_{\{n_{i}\}}^{N} &=& \frac{N!}{\prod\limits_{i=1}^{W} n_{i}!} \ ,
\end{eqnarray}
where $W$ is the total number of the one-particle states of the system, $C_{\{n_{i}\}}^{N}$ is the multinomial coefficient that defines the  Maxwell-Boltzmann statistics of particles, $\{n_{i}\}$ is the set of occupation numbers,  $\langle n_{i}\rangle$ is the mean occupation number for $i$-th state with a single-particle energy $\varepsilon_{i}$, and $p_{i}$ is the single-particle probability (the Maxwell-Boltzmann distribution). For the Maxwell-Boltzmann statistics of particles in the canonical and grand canonical ensembles, we have $p_{i}=x_{i}/\omega$, where $x_{i}=e^{-\varepsilon_{i}/T}$ and $\omega=\sum_{i=1}^{W}x_{i}$. Here, the variable $T$ is the temperature and we use a system of units in which $k_{B}=\hbar=c=1$, where $k_{B}$ is Boltzmann's constant. For instance, the Boltzmann-Gibbs entropy (\ref{1}) for the classical ideal gas in the approximation of the degenerate single-particle states of the system, i.e., the energy levels are $\varepsilon_{i}=\varepsilon$ for all states
$i=1,\ldots,W$, in the microcanonical, canonical and grand canonical ensembles is found to be
\begin{equation}\label{3a}
   S_{BG}=N\ln W, \qquad p_{i}=\frac{1}{W}.
\end{equation}
The proof for the microcanonical ensemble is given below. Thus, the equiprobable single-particle distribution (\ref{3a}) corresponds to the approximation of the degenerate single-particle states. Note that in Ref.~\cite{Oikon} the specific entropy $S_{BG}/N$ is used instead of the entropy $S_{BG}$ given in Eq.~(\ref{1}).

The Boltzmann-Gibbs entropy~(\ref{1}) is related to the classical unphysical ideal gas without the Gibbs correcting term $N!$. Therefore, the Boltzmann-Gibbs entropy for the ideal Maxwell-Boltzmann gas with the "correct Boltzmann counting" of microstates~\cite{Huang} can be written as
 \begin{eqnarray}\label{4}
    S_{BG} = N - N\ln N - N\sum\limits_{i=1}^{W} p_{i} \ln p_{i} &=&
   \ln W\{\langle n_{i}\rangle\}, \\
   \label{5}
    W\{n_{i}\} = \lim_{N\to\infty} \frac{C_{\{n_{i}\}}^{N}}{N!} &=&
    \prod\limits_{i=1}^{W}
    \left(\frac{e}{n_{i}}\right)^{n_{i}}.
\end{eqnarray}
Moreover, the Boltzmann-Gibbs entropy for the Fermi-Dirac and the Bose-Einstein  statistics of particles in the grand canonical ensemble takes the form~\cite{Landau}
\begin{equation}\label{6}
    S_{BG} = -\sum\limits_{i=1}^{W} g_{i} \left[
    \frac{\langle n_{i}\rangle}{g_{i}}\ln \frac{\langle n_{i}\rangle}{g_{i}}\pm
    \left( 1\mp \frac{\langle n_{i}\rangle}{g_{i}} \right)
\ln \left( 1\mp \frac{\langle n_{i}\rangle}{g_{i}} \right)
    \right] = \ln W\{\langle n_{i}\rangle\},
\end{equation}
where
\begin{eqnarray}\label{7}
    W\{n_{i}\}&=& \lim_{n_{i}\to\infty} C{\{n_{i}\}}, \\ \label{8}
  C{\{n_{i}\}} &=& \prod\limits_{i=1}^{W}\frac{g_{i}!}{n_{i}!(g_{i}-n_{i})!} \qquad (\mathrm{Fermi}), \qquad
   C{\{n_{i}\}} =  \prod\limits_{i=1}^{W}\frac{(n_{i}+g_{i}-1)!}{n_{i}!(g_{i}-1)!} \qquad (\mathrm{Bose}).
\end{eqnarray}
Here the upper and lower signs correspond to the Fermi and Bose statistics, respectively, and $g_{i}$ is the number of degenerate sublevels for $i$-th single-particle state. Note that in Eqs.~(\ref{1}), (\ref{4}) the number $g_{i}=1$. In paper~\cite{Oikon}, the Boltzmann-Gibbs entropy $S_{BG}$ is uniquely defined by Eq.~(\ref{1}). However, comparing Eqs.~(\ref{1})--(\ref{8}), we conclude that Eq.~(\ref{1}) cannot be the general definition for the Boltzmann-Gibbs entropy and it describes only the particular case of the unphysical ideal Maxwell-Boltzmann gas in the occupation number representation without the Gibbs correcting term~\cite{Huang}.

In the paper~\cite{Oikon}, Oikonomou proved that the Tsallis and R\'enyi entropies can be derived in the general form from the generalized multinomial coefficients which originate from the generalization of the multinomial coefficients (\ref{3}) for the Boltzmann-Gibbs statistics in the approximation (\ref{3a}) of the degenerate single-particle states of the system, $p_{i}=1/W$. He obtained
\begin{eqnarray}\label{13}
  S_{q}^{T} = N\sum\limits_{i=1}^{W} p_{i} \ln_{q} \left(\frac{1}{p_{i}}\right) &=&
  \frac{\ln_{q}(W_{q}(N))}{\phi(q)} =N\ln_{q}(W)  \qquad (\mathrm{Tsallis}), \\ \label{14}
  W_{q}(N) &=& e_{q}(N\phi(q)\ln_{q}(W))
\end{eqnarray}
and
\begin{eqnarray}\label{15}
  S_{r}^{R} = N\frac{1}{1-r}\ln \left(\sum\limits_{i=1}^{W} p_{i}^{r}\right) &=&
  \ln W_{r}(N) =N\ln W  \qquad (\mathrm{R\acute{e}nyi}), \\ \label{16}
  W_{r}(N) &=& W^{N}.
\end{eqnarray}
where
\begin{equation}\label{12}
    \ln_{q}(x)=\frac{x^{1-q}-1}{1-q}, \qquad e_{q}(x)=[1+(1-q)x]_{+}^{\frac{1}{1-q}}.
\end{equation}
Here, $p_{i}$ is the single-particle probability, $W$ is the total number of one-particle states of the system and $q,r\in \mathbf{R}$ are the real parameters, $q,r\in [0,\infty]$. Note that in Ref.~\cite{Oikon} the specific entropies $S_{q}^{T}/N$, $ S_{r}^{R}/N$ are used instead of the entropies $S_{q}^{T}$, $ S_{r}^{R}$ given in Eqs.~(\ref{13}) and (\ref{15}), respectively.  Since Eqs.~(\ref{13})--(\ref{16}) generalize Eqs.~(\ref{1})--(\ref{3}) in the approximation (\ref{3a}) for the single-particle distributions $p_{i}$, we conclude that  the Tsallis and R\'enyi entropies (\ref{13})--(\ref{16}) are the entropies of the unphysical ideal Maxwell-Boltzmann gas without the Gibbs correcting term in the approximation of the degenerate single-particle states of the system in the Tsallis and R\'enyi statistics.

In order to verify this, let us consider the ideal Maxwell-Boltzmann gas without the Gibbs correcting term in the microcanonical ensemble with $W$ degenerate one-particle states in the framework of the Boltzmann-Gibbs, Tsallis and R\'enyi statistics.

In the general case, the Boltzmann-Gibbs, Tsallis and R\'enyi statistical entropies are defined as~\cite{Tsal88,Renyi,Wehrl,Zub}
\begin{eqnarray}\label{9}
    S_{BG} &=& -\sum\limits_{n} f_{n}\ln f_{n}  \qquad \qquad (\mathrm{Gibbs}), \\
\label{10} S_{q}^{T} &=& \sum\limits_{n} f_{n}\ln_{q} \left(\frac{1}{f_{n}}\right) \qquad
(\mathrm{Tsallis}), \\ \label{11}
  S_{r}^{R} &=&  \frac{1}{1-r}\ln \left(\sum\limits_{n} f_{n}^{r}\right) \qquad (\mathrm{R\acute{e}nyi}),
\end{eqnarray}
where $f_{n}$ is the probability of $n$-th microstate (the phase distribution function or the statistical operator) and $\ln_{q}(x)$ is defined in Eq.~(\ref{12}). The expectation value for a dynamical variable and the norm equation for a distribution function can be written as~\cite{Zub}
\begin{equation}\label{17}
    \langle A\rangle =\sum\limits_{n} f_{n} A_{n}, \qquad \sum\limits_{n} f_{n}=1,
\end{equation}
where $A_{n}$ is the operator of the dynamical variable and the summation is taken over microstates of the system. Note that all entropies (\ref{1}), (\ref{4}) and (\ref{6}) for the ideal gas of the Boltzmann-Gibbs statistics were derived from their general definition (\ref{9}). Moreover, there exists an essential deference between formulas (\ref{13}) and (\ref{10}) for the Tsallis statistics. The entropy (\ref{13}) is the sum over single-particle states with the single-particle probabilities $p_{i}$, but the entropy (\ref{10}) is the sum over microstates of the system with the probabilities of microstates $f_{n}$.

Consider the equilibrium statistical ensemble of the closed energetically isolated systems of $N$ particles at the constant volume $V$, i.e., the microcanonical ensemble $(E,V,N)$. Suppose, these systems have identical energy $E$ within $\Delta E\ll E$; then the probability distribution for microstates $f_{n}$ is distinct from zero only in the layer $E\leq E_{n}\leq E +\Delta E$, where $E_{n}$ is the energy of $n$-th microstate. In this paper, for simplicity, we consider the parameters $q,r$, which were defined in Eqs.~(\ref{10}), (\ref{11}), to be the universal constants. Another interpretation for them~\cite{Parv1,ParvBiro1} is beyond the scope of the present paper. To express the equilibrium distribution function $f_{n}$ through the macroscopic variables of state of the microcanonical ensemble $(E,V,N)$, we use the thermodynamical method based on the fundamental equation of thermodynamics~\cite{Parv1,ParvBiro1} \begin{equation}\label{18}
    (dS)_{E,V,N}=0, \qquad \frac{\delta S}{\delta f_{n}}=\alpha
\end{equation}
or the Jaynes principle of maximum entropy~\cite{Tsal88,ParvBiro,Zub}
\begin{equation}\label{19}
    \Phi=S-\alpha \sum\limits_{n} f_{n}, \qquad \frac{\partial \Phi}{\partial f_{n}}=0,
\end{equation}
where $\Phi$ is the Lagrange function for the microcanonical ensemble and $\alpha$ is either a parameter (\ref{18}) or a Lagrange multiplier (\ref{19}). Combining Eqs.~(\ref{9})-(\ref{19}) we obtain the equiprobable distribution function for the Boltzmann-Gibbs, Tsallis and R\'enyi statistics in the microcanonical ensemble~\cite{Tsal88,Toral,Parv1,ParvBiro,ParvBiro1,Zub}
\begin{equation}\label{20}
   f_{n} = \frac{1}{\Gamma_{N}},  \qquad \Gamma_{N} = \sum\limits_{n}  \delta\left(E_{n}-E\right),
\end{equation}
where $\Gamma_{N}$ is the statistical weight or the total number of microstates of the system in the microcanonical ensemble. The detailed derivation of the microcanonical distribution function (\ref{20}) for the Boltzmann-Gibbs, Tsallis and R\'enyi statistics can be found in Ref.~\cite{Zub}, Refs.~\cite{Tsal88,Toral,Parv1} and Refs.~\cite{ParvBiro,ParvBiro1}, respectively.

The statistical weight (\ref{20}) for the ideal Maxwell-Boltzmann gas without the Gibbs correcting term in the microcanonical ensemble can be written as~\cite{Huang}
\begin{eqnarray}\label{21}
     \Gamma_{N} &=& \sum\limits_{\{n_{i}\}} C_{\{n_{i}\}}^{N} \
    \delta\left(\sum\limits_{i=1}^{W}n_{i}-N\right)  \delta\left(\sum\limits_{i=1}^{W}n_{i}\varepsilon_{i}-E\right).
\end{eqnarray}
Now in the approximation of degenerate one-particle states, i.e., $\varepsilon_{i}=\varepsilon$ for all levels $i=1,\ldots,W$, the statistical weight (\ref{21}), the mean occupation numbers $\langle n_{i}\rangle$ (\ref{17}) and the one-particle distribution function $p_{i}$ defined in Eq.~(\ref{1}) take the following form~\cite{Abram}
\begin{eqnarray}\label{22}
   \Gamma_{N}&=&\sum\limits_{\{n_{i}\}} C_{\{n_{i}\}}^{N} \
    \delta\left(\sum\limits_{i=1}^{W}n_{i}-N\right) = W^{N}= W\{\langle n_{i}\rangle\} =
    \left[\prod\limits_{i=1}^{W} p_{i}^{-p_{i}}\right]^{N}, \\
    \label{23} \nonumber
    \langle n_{i}\rangle &=& \frac{1}{\Gamma_{N}}\sum\limits_{\{n_{i}\}} n_{i} C_{\{n_{i}\}}^{N} \
    \delta\left(\sum\limits_{i=1}^{W}n_{i}-N\right)= \\
   &=& \frac{N}{\Gamma_{N}}\sum\limits_{\{n_{i}\}} C_{\{n_{i}\}}^{N-1} \
    \delta\left(\sum\limits_{i=1}^{W}n_{i}-(N-1)\right)= \frac{N \Gamma_{N-1}}{\Gamma_{N}} = \frac{N}{W}, \\
    \label{24}
    p_{i} &=& \frac{\langle n_{i}\rangle}{N}=\frac{1}{W},
\end{eqnarray}
where $W\{ n_{i}\}$ was defined in Eq.~(\ref{2}) and $\varepsilon N=E$. Then the microcanonical distribution function (\ref{20}) for the ideal gas in this approximation for all three statistics  (Boltzmann-Gibbs, Tsallis and R\'enyi) can be written as
\begin{eqnarray} \label{25}
 f_{n} &=& \frac{1}{\Gamma_{N}} = \frac{1}{W^{N}}= \frac{1}{W\{\langle n_{i}\rangle\}}=
 \left[\prod\limits_{i=1}^{W} p_{i}^{-p_{i}}\right]^{-N}.
\end{eqnarray}
The mean occupation numbers (\ref{23}) were obtained by the method used in Ref.~\cite{Parv3}. Substituting Eq.~(\ref{25}) into (\ref{9})--(\ref{11}) we obtain the Boltzmann-Gibbs, Tsallis and R\'enyi entropies for the ideal Maxwell-Boltzmann gas in the microcanonical ensemble without the Gibbs correcting term in the approximation of the degenerate single-particle states of the system, $p_{i}=1/W$,
\begin{eqnarray} \label{26}
 S_{BG}  &=&  -\sum\limits_{n=1}^{\Gamma_{N}} f_{n}\ln f_{n}= \ln \Gamma_{N} =\ln W\{\langle n_{i}\rangle\}=
 N\ln W =-N\sum\limits_{i=1}^{W} p_{i} \ln p_{i}  \qquad \qquad \qquad (\mathrm{Gibbs}), \\ \label{27}
   S_{q}^{T}  &=& \sum\limits_{n=1}^{\Gamma_{N}} f_{n}\ln_{q} \left(\frac{1}{f_{n}}\right) =
   \ln_{q}(\Gamma_{N}) =\ln_{q}(W\{\langle n_{i}\rangle\})=\ln_{q}(W^{N}) =
   \ln_{q}\left(e^{-N\sum_{i=1}^{W} p_{i} \ln p_{i}}\right) (\mathrm{Tsallis}),  \\ \label{28}
   S_{r}^{R} &=&  \frac{1}{1-r}\ln \left(\sum\limits_{n=1}^{\Gamma_{N}} f_{n}^{r}\right) =
   \ln \Gamma_{N} =\ln W\{\langle n_{i}\rangle\}=N\ln W =
   N\frac{1}{1-r}\ln \left(\sum\limits_{i=1}^{W} p_{i}^{r}\right) \qquad (\mathrm{R\acute{e}nyi}) , \;\;\;\;
\end{eqnarray}
where $S_{q}^{T}=\ln_{q}\left(e^{S_{BG}}\right)$ and $S_{r}^{R}= S_{BG}$. The entropy (\ref{26}) coincides with the Boltzmann-Gibbs entropy (\ref{1}), (\ref{3a}). This completes the proof for the microcanonical ensemble. The R\'enyi entropy (\ref{28}) as a function of the variables of state $(E,V,N)$ is equivalent to the Boltzmann-Gibbs entropy (\ref{26}) since in the microcanonical ensemble the R\'enyi statistics is equivalent to the
Boltzmann-Gibbs statistics for not only an ideal gas but also the general case. For more details, see Refs.~\cite{ParvBiro,ParvBiro1}. Moreover, the R\'enyi entropy (\ref{15}) was correctly obtained by Oikonomou only for the particular case of the ideal Maxwell-Boltzmann gas without the Gibbs correcting term in the approximation of the degenerate single-particle states of the system. The Tsallis entropy (\ref{13}) obtained by Oikonomou from the generalized multinomial coefficients does not coincide with the Tsallis entropy (\ref{27}) derived from the first principles. Hence, we have proved that the method of generalized multinomial coefficients failed to determine the Tsallis entropy and it is unfounded for the Tsallis statistics. This statement also applies to Ref.~\cite{Suyari}. 

Let us comment other mistakes of the paper~\cite{Oikon}. Oikonomou defines the distribution function $f_{n}$ for the Tsallis statistics as
\begin{equation}\label{29}
    f_{n}=\frac{1}{W_{q}(N)} =\frac{1}{e_{q}(N\phi(q)\ln_{q}(W))}.
\end{equation}
It is obvious that there is a discrepancy between the Oikonomou formula (\ref{29}) and the distribution function for the Tsallis statistics in the microcanonical ensemble (\ref{20}), (\ref{25}) obtained from the first principles. Thus, the Oikonomou definition of the distribution function for the Tsallis statistics is also unfounded.

Furthermore, in~\cite{Oikon}, the author compares the microcanonical equiprobable distribution function (\ref{29}) with the canonical distribution function $f_{n}\sim 1/e_{q}(\alpha+\beta(E_{n}-E))$ which originates from the Jaynes principle of maximum entropy given by the Lagrange function of the canonical ensemble~\cite{Tsal88,ParvBiro,Zub}
\begin{equation}\label{30}
    \Phi=S- \alpha\sum\limits_{n} f_{n} - \beta\sum\limits_{n} f_{n} E_{n},
\end{equation}
where $\alpha$, $\beta$ are the Lagrange multipliers and $E_{n}$ is the energy of microstates. It enables the author in Ref.~(\ref{1}) to draw far-reaching conclusions about the validity of the Jaynes principle of maximum entropy. However, a direct comparison between the microcanonical distribution function (\ref{29}) and the canonical one is incorrect, because they are two different kinds of physical objects which in the Boltzmann-Gibbs statistics are related through a Laplace transform. Therefore, the analysis like this one cannot be used to verify the validity of the Jaynes principle.

Oikonomou~\cite{Oikon} rewrote the equal-probability distribution function (\ref{25}) for the R\'enyi statistics in the form
\begin{equation}\label{31}
   f_{n} = \frac{1}{W_{r}(N)}, \qquad W_{r}(N)= W^{N} = e^{N\ln W}.
\end{equation}
Based on this formula, he claimed in~\cite{Oikon} that for the R\'enyi entropy $W_{r}(N)$ is given by an ordinary exponential distribution. This misleading statement suggests that the R\'enyi statistics has an exponential distribution function. It is well known that the thermodynamical self-consistent R\'enyi statistics is characterized by the power-law distribution function in the canonical ensemble and the equiprobable distribution function in the microcanonical ensemble~\cite{ParvBiro,ParvBiro1}. Evidently, the distribution function (\ref{31}) is equiprobable and it cannot be an exponential function. However, for some special cases of a generalized entropy maximization procedure the R\'enyi statistics in the canonical ensemble may be described by the exponential distribution function, too~\cite{Bagci}.

The difference between (\ref{13}) and (\ref{27}) must be a consequence of the very non-extesivity of the Tsallis entropy formula. The other two entropy formulas are known to be statistically additive. Hence, the $N$ particle ideal gas results obtained under the microcanonical constraints do agree.  

This work was supported in part by the joint research project of JINR and IFIN-HH, protocol N~3891-3-09/09, N~4006, the RFBR grant~08-02-01003-a and the MTA-JINR grant. I acknowledge valuable remarks and fruitful discussions with T.S.~Bir\'{o}. The author is also indebted to P\'eter V\'an for drawing his attention to Ref.~\cite{Renyi} on the first appearance of R\'eny entropy.

\end{document}